\documentclass[prb,preprint,showpacs,amsmath,amssymb,floatfix,prb,superscriptaddress,showkeys,aps]{revtex4-1}

\usepackage{graphicx}
\usepackage{epsfig}
\usepackage{dcolumn}
\usepackage{bm}
\usepackage{times}

\begin{document}


\title{Relation between Poisson and Schr\"odinger equation}

\author{Gabriel Gonz\'alez}
\email{gabrielglez@iteso.mx}
\affiliation{Departamento de Matem\'aticas y F\'isica, Instituto Tecnol\'ogico y de Estudios Superiores de Occidente, \\ Perif\'erico Sur Manuel G\'omez Mor{\'i}n 8585 C.P. 45604, Tlaquepaque, Jal., MEXICO}

\date{\today}

\begin{abstract}
The relation between the Poisson and Schr\"odinger equation is obtained through a simple transformation. It is pointed out that this analogy between both equations can be only applied for potentials that involve a combination of attractive and repulsive delta function potentials. This relationship enables us to use elementary electrostatic results to find the ground state energy and wave function of the associated quantum state problem in one dimension. Particularly, the result shows that it is possible to trap a single electron in a one dimensional ionic crystal modeled by repulsive and attractive delta functions.
\end{abstract}



\maketitle

\section{Introduction}
\label{sec:introduction}
It is very well known that Poisson's equation is among the most important equations in electromagnetic theory. This equation fundamentally links the source of the field, i.e. charge distribution $\rho$, and the electric potential $V$. In theory, this equation allow us to find all we need to know about an electrostatic field given the charge distribution. The general solution to Poisson's equation has the following integral form \cite{d1}
\begin{equation}
V(\vec{r})=\frac{1}{4\pi\epsilon_0}\int\frac{\rho(\vec{r}^{\,\prime})}{|\vec{r}-\vec{r}^{\,\prime}|}d^{3}\vec{r}^{\,\prime}.
\label{eqi1}
\end{equation}
Once we know $V$ the electric field is calculated from the gradient of $V$, i.e. $\vec{E}=-\nabla V$. Nevertheless, the integral given in Eq. (\ref{eqi1}) can only be evaluated analytically for the simplest charge configurations.
This is why it is very useful to recast the problem using Poisson's equation with appropiate boundary conditions. For the restricted case of static charges and fields in vacuum, the appropiate boundary conditions for the electric potential are \cite{stew}
\begin{equation}
\left\{ \begin{array}{ll}
V & \mbox{is always a continuous function across any boundary, and}\\ 
\Delta(dV/dn)=-\sigma/\epsilon_0 & \mbox{the normal derivative of the potential is discontinuous at the boundary,}\\ & \mbox{where $\sigma$ represents the surface charge distribution at the boundary.} \end{array} \right.
\label{eqi1a}
\end{equation}
On the other hand, the standard boundary conditions for the wave function $\psi$ in one dimensional quantum mechanics are :\cite{d2}
\begin{equation}
\left\{ \begin{array}{ll}
																				\psi & \mbox{is always continuous, and}\\ 
																				d\psi/dz & \mbox{is continuous except at points where the potential is infinite.}
																				\end{array} \right.
\label{eqi2}
\end{equation}
In particular, if there is a delta function potential at $z_0$, then the change of slope of the wave function, $\Delta(d\psi/dz)$, at the point $z_0$ is proportional to the amplitude of the delta function as well as the wave function $\psi(z_0)$ at this point.\cite{ma} This is the type of boundary conditions we expect for the electrostatic potential. As a consequence, the electrostatic potential and the one dimensional wave function share the same boundary conditions for a delta function potential.\\ The one dimensional delta-function potential has proved to be a useful model to represent diatomic ion and molecular orbitals,\cite{am1} two electron systems,\cite{am2} electron-dipole interaction,\cite{am3} solids,\cite{am4} the many body problem,\cite{am5} and scattering.\cite{am6} It is well known that the single attractive delta function potential, i.e. $U(z)=-\alpha\delta(z)$, possesses one exponentially localized bound state for all values of the strength of the well $\alpha>0$. Its existence and stability has been tested for the effects of different boundary conditions\cite{ajp1} and symmetry-breaking perturbations.\cite{ajp2}\\
In this article, a mapping between a three dimensional electrostatic problem described by Poisson's equation and the one dimensional Schr\"ondinger equation is given through a simple transformation. This relation will allow us to find the wave function and energy of the ground state of the quantum system by using electrostatic results. \\ The article is organized as follows. First we will start by transforming the Poisson equation into a Schr\"odinger-like equation, and discuss when both equations are equivalent. Then we will apply electrostatic results to solve the Schr\"odinger equation for the case of a potential function given by a combination of attractive and repulsive delta functions. The conclusions are summarized in the last section.
\section{Duality between Poisson and Schr\"odinger equation}
\label{sec:model}
The Poisson equation in three dimensions is given by 
\begin{equation}
\nabla^2V=-\frac{\rho}{\epsilon_0},
\label{eq01}
\end{equation}
where $V$ is the electric potential, $\rho$ is the volume charge density and $\epsilon_0$ is the electric permitivity of free space.\cite{d1,stew} \\ Making the following transformation $V(\vec{r})=V_0\ln(\Psi(\vec{r})/A)$, where $V_0$ and $A$ are constants to ensure dimensional consistency, we obtain
\begin{equation}
V_0\frac{\nabla^2\Psi}{\Psi}-\frac{1}{V_0}\vec{E}\cdot\vec{E}=-\frac{\rho}{\epsilon_0},
\label{eq02}
\end{equation} 
where we have used the fact that $\vec{E}=-\nabla V=-V_0\frac{\Psi^{\prime}}{\Psi}$. Multiplying Eq. (\ref{eq02}) by $V_0\epsilon_0\Psi/2$ we end up with
\begin{equation}
-\frac{V_0^2\epsilon_0}{2}\nabla^2\Psi-\frac{1}{2}V_0\rho\Psi=-\frac{\epsilon_0}{2}\vec{E}\cdot\vec{E}\Psi.
\label{eq02a}
\end{equation}
Equation (\ref{eq02a}) is a Schr\"odinger-type equation when the magnitude $|\vec{E}|$ is constant. We know from basic electrodynamic courses that the only charge distribution to produce uniform electrostatic fields are infinite charge sheets. Since the infinite charge sheet has plane symmetry, the electrostatic potential depends on one variable only, i.e. $\Psi=\psi(z)$. For this case Eq. (\ref{eq02a}) is given by 
\begin{equation}
-\frac{V_0^2\epsilon_0}{2}\frac{d^2\psi}{dz^2}-\frac{1}{2}V_0\rho\psi=-\frac{\epsilon_0}{2}E_z^2\psi.
\label{eq02b}
\end{equation} 
Making the following substitutions:
\begin{equation}
V_0^2\epsilon_0a_0^3=\frac{\hbar^2}{m}, \quad -\frac{1}{2}V_0\rho(z)a_0^3+\frac{\epsilon_0}{2}(E_z^2-E_{z\infty}^2)a_0^3=U(z) \quad\mbox{and}\quad \frac{1}{2}\epsilon_0E_{z\infty}^2a_0^3=|E|
\label{eq03}
\end{equation}
where $a_0$ is an arbitrary length and $E_{z\infty}$ is the asymptotic magnitude of the electric field, i.e. $E_{z\infty}=\lim_{z\rightarrow\infty}|E_z|$, we obtain the time independent Schr\"odinger equation 
\begin{equation}
-\frac{\hbar^2}{2m}\frac{d^2\psi}{dz^2}+U(z)\psi=-|E|\psi,
\label{eq04}
\end{equation}
for the case of bound states, i.e. $E<0$.\cite{lau} This relationship means that the wave function and the bound state energy will be given in terms of the electric potential and the energy density by
\begin{equation}
\psi(z)=Ae^{V(z)/V_0}, \quad E=-{\cal U},
\label{eq04a}
\end{equation}
where $A$ is the normalization constant and ${\cal U}=\epsilon_0 E_z^2/2$ is the energy density, respectively. Note that the wave function has no nodes, therefore only the ground state can be found with this method. This is consistent with the electrostatic uniqueness theorem which guarantees that the electrostatic potential is uniquely determined for a given charge density distribution and the electrostatic potential at all boundaries are specified.\cite{d1}\\
In order to have physically realizable quantum states the wave function must be normalized.\cite{d2,ma,lau,merz,shan}\\
In the following section we solve the Schr\"odinger equation using electrostatic solutions.

\section{Solving the Schr\"odinger equation via electrostatic solutions}
\label{sec:CQ}
\begin{figure}[ht]
  \begin{center}
    \begin{tabular}{cc}
      \resizebox{70mm}{!}{\includegraphics{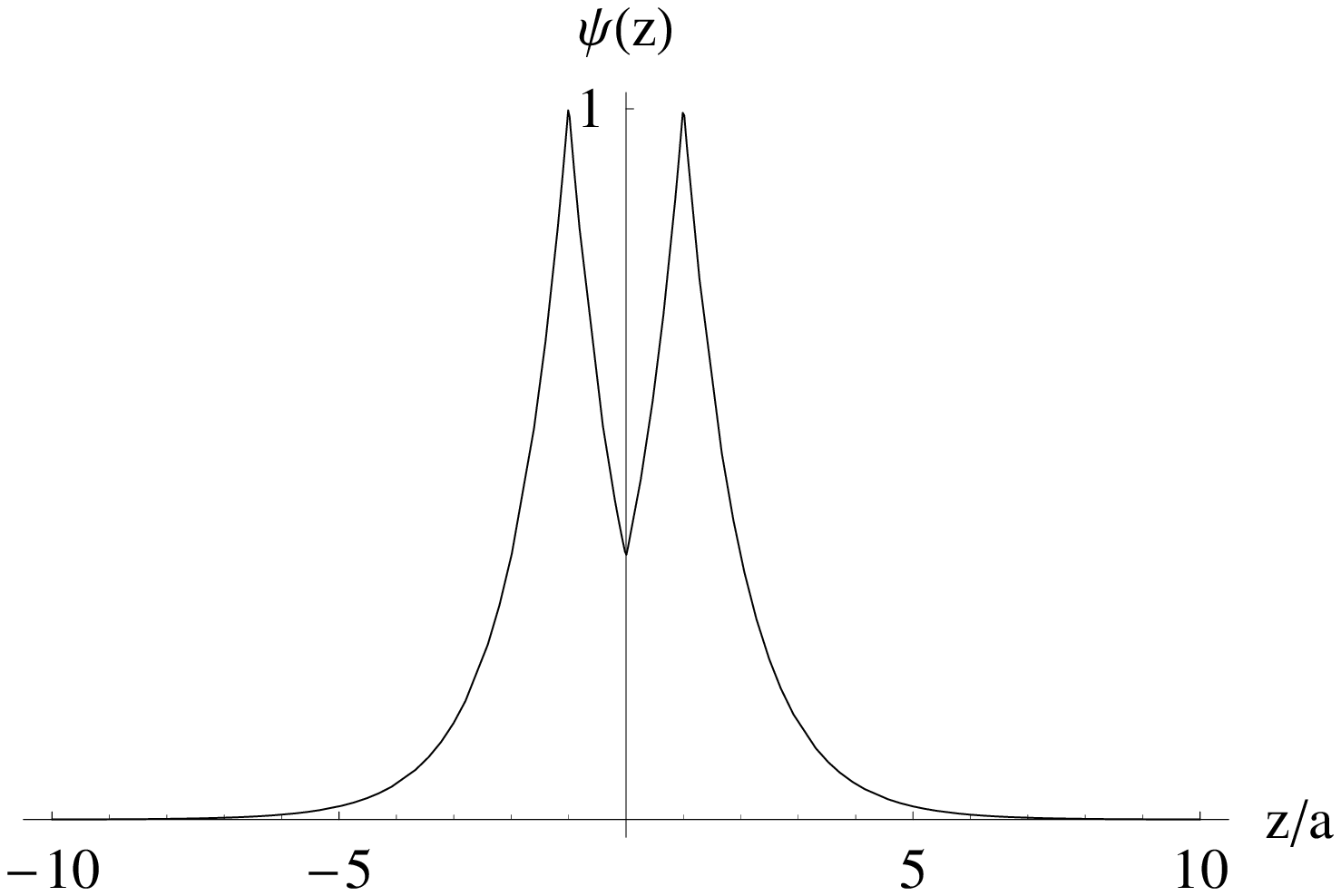}} &
      \resizebox{70mm}{!}{\includegraphics{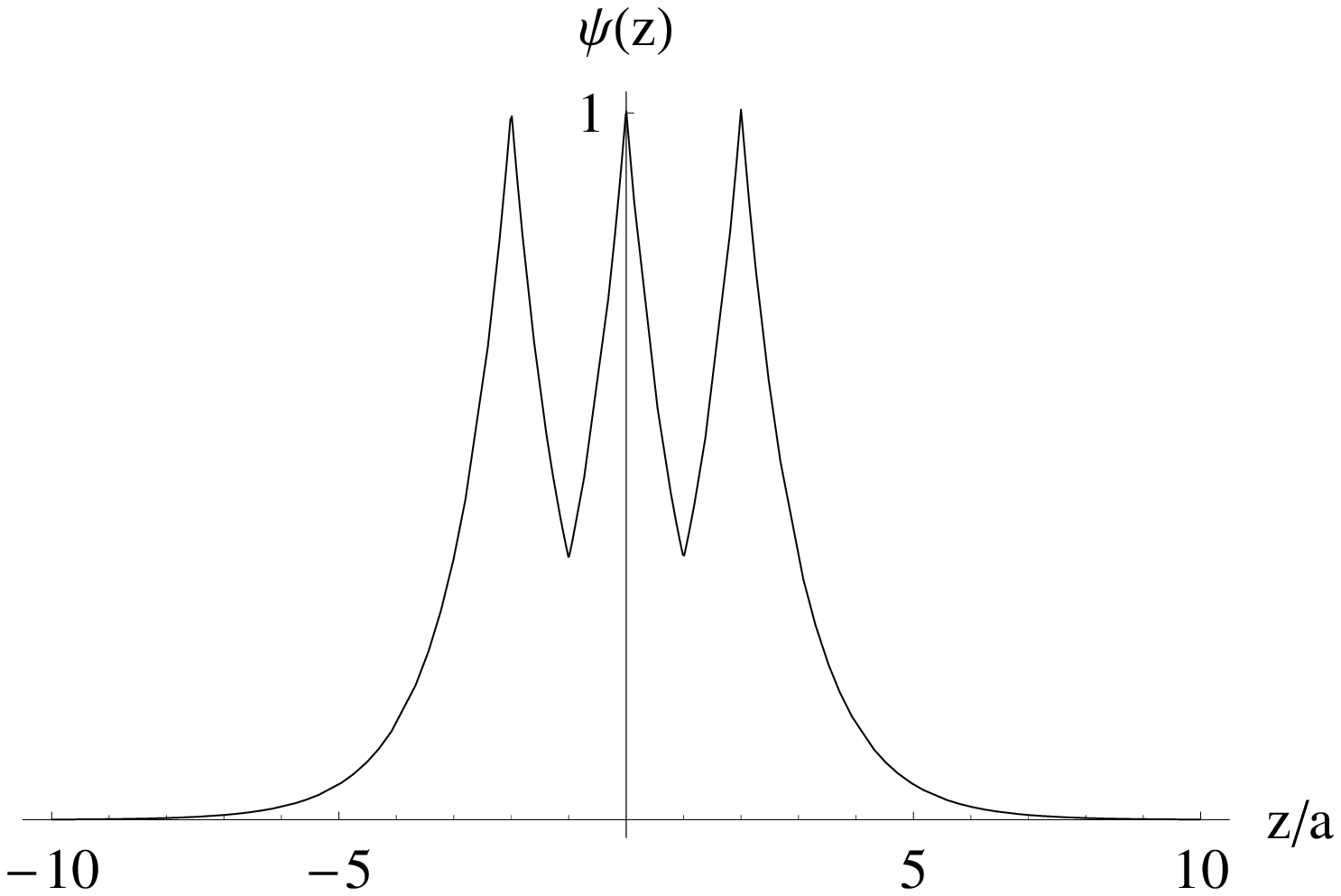}} \\ 
    \multicolumn{1}{c}{\mbox{\bf (a)}} &
		\multicolumn{1}{c}{\mbox{\bf (b)}} \\ 
      \resizebox{70mm}{!}{\includegraphics{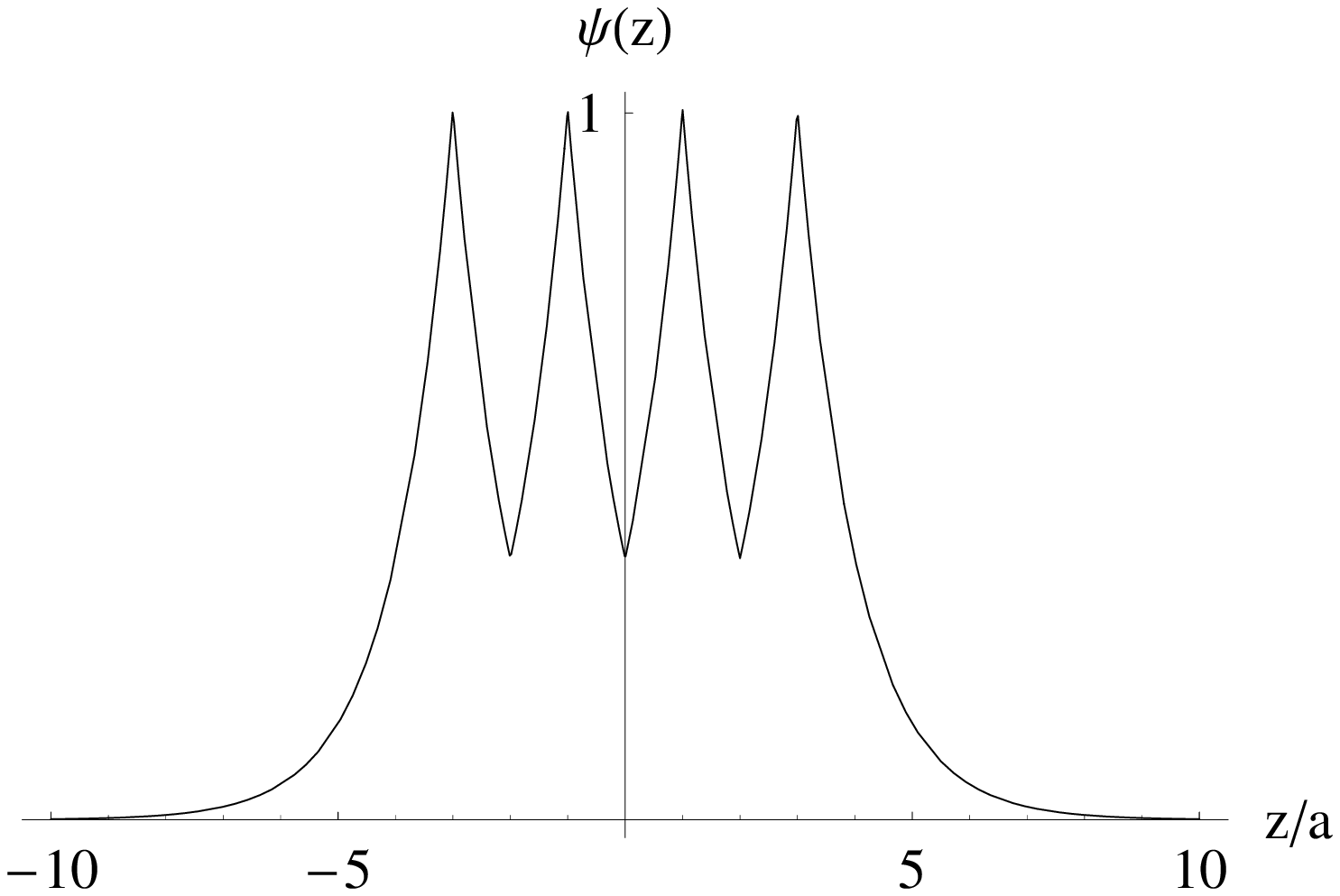}} &
      \resizebox{70mm}{!}{\includegraphics{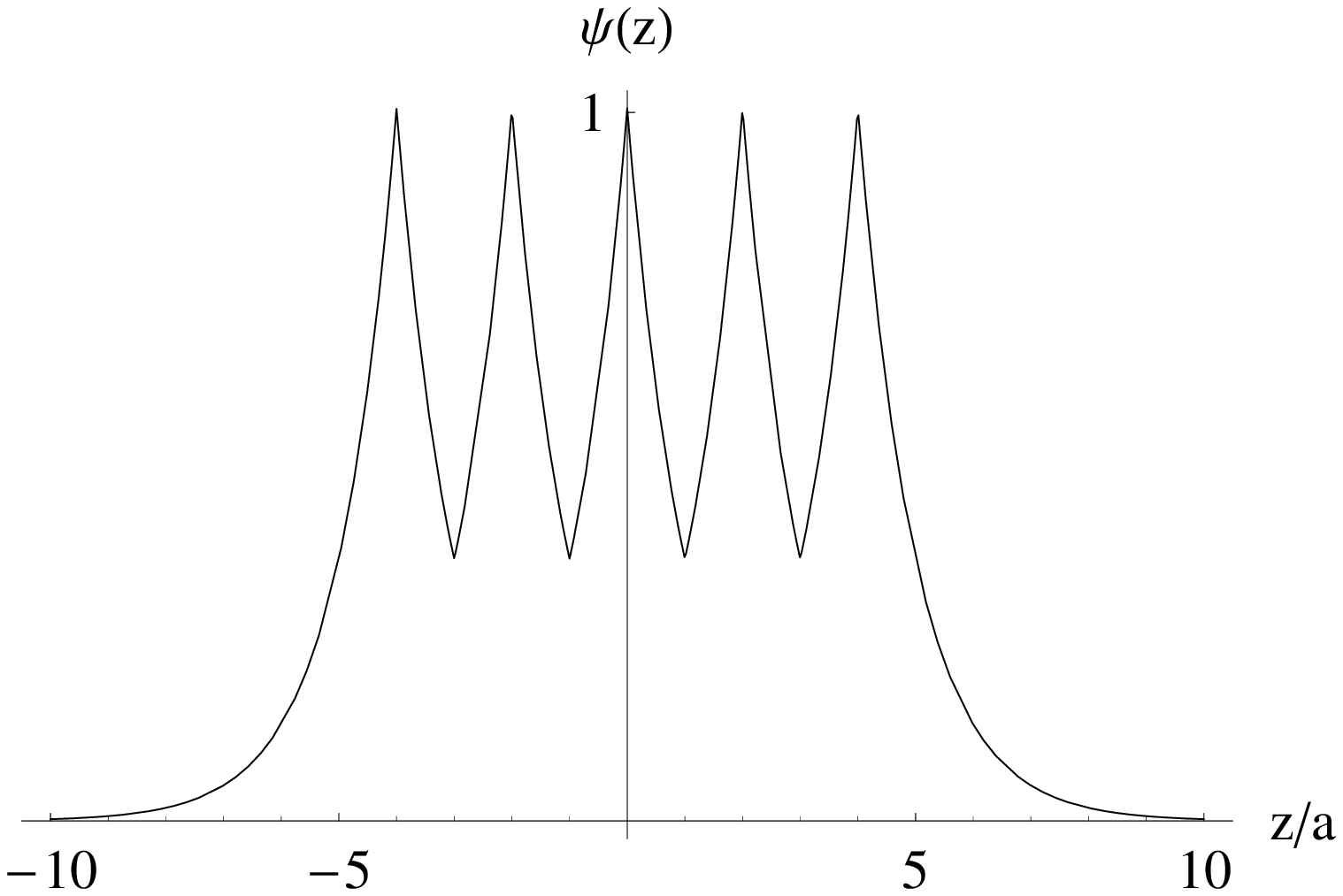}} \\
    \multicolumn{1}{c}{\mbox{\bf (c)}} &
		\multicolumn{1}{c}{\mbox{\bf (d)}} \\
    \end{tabular}
    \caption{Plots showing the wave function for the following 
    values: (a) $N=1$, (b)
    $N=2$, (c) $N=3$
    and (d) $N=4$. We used atomic units, i.e. $\hbar=m=1$, and $\alpha a=1$.}
	\label{psi}
  \end{center}
\end{figure}

We know from basic electrodynamic courses that the only charge distribution to produce a uniform electrostatic field are an odd number of parallel infinite sheets with opposite electric charge densities $\pm\sigma$ \cite{d1,stew}, hence we will consider a volume charge density given by 
\begin{equation}
\rho(z)=\sigma\sum_{n=-N}^{N}(-1)^{n}\delta(z-z_n),\,\, \mbox{for $N=0,1,2,\ldots$}
\label{eq05}
\end{equation}
where $\delta(z-z_n)$ is the Dirac delta function at the position $z_n$.\cite{hassani} 
Let us consider the case when $N=0$, for this case the volume charge distribution is of the form
\begin{equation}
\rho(z)=\sigma\delta(z),
\label{eq06}
\end{equation}
where we have taken $z_0=0$. The electric field caused by an infinite sheet of charge density $\sigma$ at any space point is \cite{d1,stew}
\begin{equation}
\vec{E}_z=\frac{\sigma }{2\epsilon_0}\frac{z}{|z|}\hat{e}_z,
\label{eq07}
\end{equation}
therefore, the electric potential and the energy density are given by \cite{d1,stew}
\begin{equation}
V(z)=-\frac{\sigma}{2\epsilon_0}|z|, \quad {\cal U}=\frac{\sigma^2}{8\epsilon_0}.
\label{eq08}
\end{equation}
Using equation (\ref{eq04a}) we obtain the wave function and the energy of the bound state
\begin{equation}
\psi(z)=Ae^{-\sigma|z|/2\epsilon_0V_0}, \quad E=-\frac{\sigma^2}{8\epsilon_0}. 
\label{eq09}
\end{equation}
The wave function given by equation (\ref{eq09}) can be normalized only for $\sigma>0$, otherwise it blows up when $z\rightarrow\pm\infty$.
Using equation (\ref{eq03}) we get the potential energy function $U(z)=-\alpha\delta(z)$ for the quantum mechanical case, where $\alpha=\sigma V_0/2$ is the strength of the potential. Making the appropiate substitutions into equation (\ref{eq09}) we get the corresponding bound state solution in the presence of an atractive delta function potential
\begin{equation}
\psi(z)=Ae^{-m\alpha|z|/\hbar^2}, \quad E=-\frac{m\alpha^2}{2\hbar^2},
\label{eq10}
\end{equation}
where the normalization constant is $A=\sqrt{m\alpha}/\hbar$.\cite{d2} Using the wave function given in equation (\ref{eq10}) it is easy to obtain the expectation values for the potential energy $\langle U\rangle_0=-m\alpha^2/\hbar^2$ and kinetic energy $\langle T\rangle_0=m\alpha^2/2\hbar^2$, respectively. \\
If we consider the general case for $N\geq1$, then the electrostatic potential and energy density are given by
\begin{equation}
V(z)=-\frac{\sigma}{2\epsilon_0}\sum_{n=-N}^{N}(-1)^{n}|z-z_n|, \quad {\cal U}=\frac{\sigma^2}{8\epsilon_0}.
\label{eq11}
\end{equation}
Interestingly, there is one quantum bound state with exactly the same energy for the charge distributions given by equations (\ref{eq05}) and (\ref{eq06}), where the wave function is given by
\begin{equation}
\psi(z)=A\exp\left(-\frac{\sigma}{2\epsilon_0V_0}\sum_{n=-N}^{N}(-1)^n|z-z_n|\right).
\label{eq12}
\end{equation}
Remarkably, the solution given in equation (\ref{eq12}) is the product of growing and decaying exponential functions. 
In order to have a physically realizable quantum bound states we need to normalize the wave function given in equation (\ref{eq12}). For simplicity, I am going to consider the case consisting of evenly spaced infinite charge sheets, i.e. $z_n=na$, where  $a$ is the distance between adjacent sheets. For this particular case the wave function is an even function and in order to normalize $\psi$ we need to solve the following integral
\begin{equation}
2|A|^2\int_{0}^{\infty}\exp\left(-\frac{\sigma}{\epsilon_0V_0}\sum_{n=-N}^{N}(-1)^n|z-na|\right)dz=1.
\label{eq13}
\end{equation}
Using the fact that
\begin{equation}
\sum_{n=-N}^{N}(-1)^n|z-na|=\sum_{n=-N}^{0}(-1)^n|z-na|+\sum_{n=1}^{N}(-1)^n|z-na|,
\label{eq14}
\end{equation}
and by changing $n$ by $-n$ in the first sum we have
\begin{equation}
\sum_{n=-N}^{N}|z-na|=\sum_{n=0}^{N}|z+na|+\sum_{n=1}^{N}|z-na|,
\label{eq14}
\end{equation}
then the integral given in equation (\ref{eq13}) reduces to
\begin{widetext}
\begin{eqnarray}
 \int_{0}^{Na}\exp\left[-\frac{\sigma}{\epsilon_0V_0}\left(\sum_{n=0}^{N}(-1)^n|z+na|+\sum_{n=1}^{N}(-1)^n|z-na|\right)\right]dz+\\ \nonumber \int_{Na}^{\infty}\exp\left[-\frac{\sigma}{\epsilon_0V_0}\left(\sum_{n=0}^{N}(-1)^n|z+na|+\sum_{n=1}^{N}(-1)^n|z-na|\right)\right]dz.
\label{eq15}
\end{eqnarray}
\end{widetext}
The result of the integral given above is (the evaluation of this integral is given in appendix \ref{appendixA})
\begin{widetext}
\begin{equation}
\frac{1}{2|A|^2}=\frac{2\epsilon_0V_0}{\sigma}\exp\left[-\frac{(1+2N)\sigma a(-1)^N}{2\epsilon_0V_0}\right]\sum_{k=0}^{N-1}\sinh\left[(-1)^k\frac{\sigma a}{2\epsilon_0V_0}\right]-\frac{\epsilon_0V_0}{\sigma(-1)^N}\exp\left[-\frac{\sigma(-1)^N}{\epsilon_0V_0}z\right]\Bigg|_{Na}^{\infty}
\label{eq16}
\end{equation}
\end{widetext}
The result given in equation (\ref{eq16}) will diverge unless $\sigma\rightarrow(-1)^{N}\sigma$, for $\sigma>0$. This restriction means that we need to have positive infinite charge sheets on both ends of the configuration in order for the wave function to go to zero when $z\rightarrow\pm\infty$. Therefore, the normalization constant is given by
\begin{widetext}
\begin{equation}
A^{-1}=\sqrt{\frac{2\epsilon_0V_0}{\sigma}\exp\left[-\frac{\sigma Na}{\epsilon_0V_0}\right]+\frac{4\epsilon_0V_0}{\sigma}\exp\left[-\frac{(1+2N)\sigma a}{2\epsilon_0V_0}\right]\left[\frac{1-(-1)^N}{2}\right]\sinh\left[\frac{\sigma a}{2\epsilon_0V_0}\right]}\,\,\mbox{for $N\geq1$.}
\label{eq17}
\end{equation}
\end{widetext}
where we have used the fact that
\begin{equation}
(-1)^N\sum_{k=0}^{N-1}\sinh\left[(-1)^{k+N}\frac{\sigma a}{2\epsilon_0V_0}\right]=\left[\frac{1-(-1)^N}{2}\right]\sinh\left[\frac{\sigma a}{2\epsilon_0V_0}\right].
\label{eq18}
\end{equation}
Therefore, if we make the appropiate substitutions we have the bound state energy and wave function for a quantum particle of mass $m$ moving through the potential energy function of a one-dimensional ionic crystal \cite{hassani}, i.e.
\begin{equation}U(z)=-\alpha\sum_{n=-N}^{N}(-1)^{n+N}\delta(z-na),
\label{eq18a}
\end{equation} 
which is given by
\begin{widetext}
\begin{equation}
\psi(z)=\frac{{\displaystyle\prod_{n=-N}^{N}}\exp\left[-\frac{\alpha m(-1)^{n+N}}{\hbar^2}|z-na|\right]}{\sqrt{\frac{\hbar^2}{m\alpha}\exp\left[-\frac{2m\alpha Na}{\hbar^2 }\right]+\frac{2\hbar^2 }{m\alpha}\exp\left[-\frac{(1+2N)am\alpha }{\hbar^2}\right]\left[\frac{1-(-1)^N}{2}\right]\sinh\left[\frac{m\alpha a}{\hbar^2}\right]}},
\label{eq19}
\end{equation}
\end{widetext} 
for $N\geq1$ and with energy $E=-\frac{m\alpha^2}{2\hbar^2}$.
This result shows that it is possible to trap a single electron with a one dimensional ionic crystal modeled by repulsive and attractive delta functions for the negative and positive ions, respectively. The wave function $\psi(z)$ is plotted in Fig. (\ref{psi}) for different values of $N$. 
With the wave function given by equation (\ref{eq19}) we can calculate the expectation value of the potential energy, which is given by
\begin{widetext}
\begin{equation}
\langle U\rangle=-\alpha\sum_{n=-N}^{N}(-1)^{n+N}|A|^2\int_{-\infty}^{\infty}\exp\left[-\frac{2\alpha m}{\hbar^2}\sum_{j=-N}^{N}(-1)^{j+N}|z-ja|\right]\delta(z-na)dz.
\label{eq20}
\end{equation}
\end{widetext}
solving the integral given above (the evaluation of this integral is given in appendix \ref{appendixB}), we find the expectation value of the potential energy function
\begin{equation}
\langle U\rangle=-\frac{m\alpha^2}{\hbar^2}\left(\frac{1+2Ne^{-m\alpha a/\hbar^2}\sinh[m\alpha a/\hbar^2]}{1+2e^{-m\alpha a/\hbar^2}\left[\frac{1-(-1)^N}{2}\right]\sinh[m\alpha a/\hbar^2]}\right)
\label{eq20}
\end{equation}
and the average kinetic energy which is given by
\begin{equation}
\langle T\rangle=-\frac{m\alpha^2}{\hbar^2}\left(\frac{1}{2}-\frac{1+2Ne^{-m\alpha a/\hbar^2}\sinh[m\alpha a/\hbar^2]}{1+2e^{-m\alpha a/\hbar^2}\left[\frac{1-(-1)^N}{2}\right]\sinh[m\alpha a/\hbar^2]}\right).
\label{eq21}
\end{equation}
Now we will consider the case when the volume charge density is given by
\begin{equation}
\rho=\sigma\left(\delta(z-a)+\delta(z+a)\right),
\label{eq22}
\end{equation}
the electric field for this charge configuration is given by
\begin{equation}
\vec{E}_z=\left\{ \begin{array}{ll}
																				\frac{\sigma}{\epsilon_0}\hat{e}_z & \mbox{if $z>a$}\\
																				0 & \mbox{if $-a<z<a$}\\
																				-\frac{\sigma}{\epsilon_0}\hat{e}_z & \mbox{if $z<-a$},
																				\end{array} \right.
\label{eq23}
\end{equation}
and the electrostatic potential will be
\begin{equation}
V(z)=\left\{ \begin{array}{ll}
																				-\frac{\sigma}{\epsilon_0}z & \mbox{if $z>a$}\\
																				-\frac{\sigma a}{\epsilon_0} & \mbox{if $-a<z<a$}\\
																				\frac{\sigma}{\epsilon_0}z & \mbox{if $z<-a$},
																				\end{array} \right.
\label{eq23}
\end{equation}
therefore, the dual quantum problem will have a ground state energy given by $E=-2m\alpha^2/\hbar^2$ with the following wave function
\begin{equation}
\psi(z)=\left\{ \begin{array}{ll}
																				Ae^{-2m\alpha z/\hbar^2} & \mbox{if $z>a$}\\
																				Ae^{-2m\alpha a/\hbar^2} & \mbox{if $-a<z<a$}\\
																			  Ae^{2m\alpha z/\hbar^2} & \mbox{if $z<-a$},
																				\end{array} \right.
\label{eq24}
\end{equation}
for the following quantum potential
\begin{equation}
U(z)=-\sigma\left(\delta(z-a)+\delta(z+a)\right)-\frac{2m\alpha^2}{\hbar^2}\theta(|z|<a),
\label{eq25}
\end{equation}
where $\theta(|z|<a)$ is one for $-a<z<a$ and zero otherwise. Notice that the origin of the last term in Eq. (\ref{eq25}) is due to the difference in electromagnetic energy between the asymptotic region and all space points.  
\section{Conclusions} 
We have shown a relation between Poisson and Sch\"odinger equation which enable us to find the ground state energy and eigenfunction of a dual quantum system using electrostatic solutions.
We have obtained the general solution of the wave function for a quantum particle moving through a potential energy function that describes a one-dimensional ionic crystal of finite length using electrostatic solutions. The result shows that forces exerted by regularly spaced positively and negatively ions described by attractive and repulsive Dirac delta functions admit only one bound state, i.e. it is possible to trap a single electron in a ionic crystal. The method can be used for non periodic potentials which can describe one dimensional ionic quasicrystals.

\section{Acknowledgments}

I would like to thank an anonymous referee for useful comments and feedback to improve this article.

\appendix
\section{Derivation of the normalization constant}
\label{appendixA}

We need to evaluate the integral
\begin{equation}
\int_{0}^{\infty}e^{f(z)}dz=\int_{0}^{Na}e^{f(z)}dz+\int_{Na}^{\infty}e^{f(z)}dz,
\label{eqa0}
\end{equation}
where
\begin{equation}
f(z)=-\frac{\sigma}{\epsilon_0V_0}\left(\sum_{n=0}^{N}(-1)^n|z+na|+\sum_{n=1}^{N}(-1)^n|z-na|\right).
\label{eqa1}
\end{equation}
Using the fact that 
\begin{equation}
f(z)=-\frac{\sigma}{\epsilon_0V_0}\left[(-1)^kz+\sum_{n=k+1}^{N}(-1)^{n}(2na)\right], \quad\mbox{for}\quad ka<z<(k+1)a,
\label{eqa3}
\end{equation}
then we can rewrite the first integral in equation (\ref{eqa0}) in the form
\begin{equation}
\int_{0}^{Na}e^{f(z)}dz=\sum_{k=0}^{N-1}\int_{ka}^{(k+1)a}\exp\left[-\frac{\sigma}{\epsilon_0V_0}\left[(-1)^kz+\sum_{n=k+1}^{N}(-1)^{n}(2na)\right]\right]dz.
\label{eqa4}
\end{equation}
Integrating equation (\ref{eqa4}) we get
\begin{equation}
\frac{\epsilon_0V_0}{\sigma}\sum_{k=0}^{N-1}(-1)^k\exp\left[-\frac{\sigma}{\epsilon_0V_0}\sum_{n=k+1}^{N}(-1)^{n}(2na)\right]\exp\left[-\frac{\sigma a (-1)^kk}{\epsilon_0V_0}\right]\left[1-\exp\left[-\frac{\sigma a(-1)^k}{\epsilon_0V_0}\right]\right].
\label{eqa5}
\end{equation}
Evaluating the sum over $n$ and regrouping terms in $k$ we end up with
\begin{equation}
\int_{0}^{Na}e^{f(z)}dz=\frac{2\epsilon_0V_0}{\sigma}\exp\left[-\frac{\sigma a (-1)^N(1+2N)}{2\epsilon_0V_0}\right]\sum_{k=0}^{N-1}\sinh\left[\frac{\sigma a (-1)^k}{2\epsilon_0V_0}\right].
\label{eqa6}
\end{equation}
In orde to evaluate the second integral in equation (\ref{eqa0}), we use the fact that
\begin{equation}
f(z)=-\frac{\sigma}{\epsilon_0V_0}\left[z+\sum_{n=1}^{N}(-1)^{n}(2z)\right], \quad\mbox{for}\quad Na<z<\infty,
\label{eqa7}
\end{equation}
therefore, evaluating the sum over $n$ we end up with
\begin{equation}
\int_{Na}^{\infty}e^{f(z)}dz=\int_{Na}^{\infty}\exp\left[-\frac{\sigma(-1)^N}{\epsilon_0V_0}z\right]dz,
\label{eqa8}
\end{equation}
integrating equation (\ref{eqa8}) we have
\begin{equation}
\int_{Na}^{\infty}e^{f(z)}dz=-\frac{\epsilon_0V_0}{\sigma (-1)^N}\exp\left[-\frac{\sigma(-1)^N}{\epsilon_0V_0}z\right]\Bigg|_{Na}^{\infty}.
\label{eqa9}
\end{equation}
This integral given in equation (\ref{eqa9}) clearly diverges unless we choose $\sigma=(-1)^N\sigma$, with $\sigma>0$. 
\section{Derivation of the potential energy expectation value}
\label{appendixB}
We want to calculate the expectation value of the potential energy, which is given by
\begin{equation}
\langle U\rangle=-\alpha\sum_{n=-N}^{N}(-1)^{n+N}|A|^2\int_{-\infty}^{\infty}\exp\left[-\frac{2\alpha m}{\hbar^2}\sum_{j=-N}^{N}(-1)^{j+N}|z-ja|\right]\delta(z-na)dz,
\label{eqb1}
\end{equation}
where $|A|$ is the normalization constant. Integrating equation (\ref{eqb1}) we have
\begin{equation}
\langle U\rangle=-\alpha|A|^2\sum_{n=-N}^{N}(-1)^{n+N}\exp\left[-\frac{2ma\alpha}{\hbar^2}\sum_{j=-N}^{N}(-1)^{j+N}|n-j|\right],
\label{eqb2}
\end{equation}
by using the following identity
\begin{equation}
\sum_{j=-N}^{N}(-1)^{j+N}|n-j|=\sum_{j=-N}^{n-1}(-1)^{j+N}(n-j)+\sum_{j=n+1}^{N}(-1)^{j+N}(j-n),
\label{eqb3}
\end{equation}
and if we let $n-j=k$ and $j-n=l$ in the first and second sum, respectively. These substitutions change the limits of the sums, and we get
\begin{equation}
\sum_{j=-N}^{N}(-1)^{j+N}|n-j|=\sum_{k=n+1}^{1}(-1)^{n-k+N}k+\sum_{l=1}^{N-n}(-1)^{l+n+N}l.
\label{eqb4}
\end{equation}
Evaluating the sums over $k$ and $l$ in equation (\ref{eqb4}) we get 
\begin{equation}
\sum_{j=-N}^{N}(-1)^{j+N}|n-j|=\frac{1}{2}\left[(1+2N)-(-1)^{n+N}\right].
\label{eqb5}
\end{equation}
Substituting equation (\ref{eqb5}) into equation (\ref{eqb2}) we have
\begin{equation}
\langle U\rangle=-\alpha |A|^2\exp\left[-\frac{ma\alpha(1+2N)}{\hbar^2}\right]\sum_{n=-N}^{N}(-1)^{n+N}\exp\left[\frac{ma\alpha}{\hbar^2}(-1)^{n+N}\right].
\label{eqb6}
\end{equation}
Using the fact that
\begin{equation}
\sum_{n=-N}^{N}(-1)^{n+N}\exp\left[\frac{ma\alpha}{\hbar^2}(-1)^{n+N}\right]=\sum_{n=0}^{2N}(-1)^{n}\exp\left[\frac{ma\alpha}{\hbar^2}(-1)^{n}\right]=e^{ma\alpha/\hbar^2}+2N\sinh[ma\alpha/\hbar^2],
\label{eqb7}
\end{equation}
and substituting the normalization constant, we end up with
\begin{equation}
\langle U\rangle=-\frac{m\alpha^2\exp[-\frac{\alpha m a}{\hbar^2}(1+2N)]\left(e^{ma\alpha/\hbar^2}+2N\sinh[ma\alpha/\hbar^2]\right)}{\hbar^2\exp\left[-\frac{2m\alpha Na}{\hbar^2 }\right]+2\hbar^2 \exp\left[-\frac{(1+2N)am\alpha }{\hbar^2}\right]\left[\frac{1-(-1)^N}{2}\right]\sinh\left[\frac{m\alpha a}{\hbar^2}\right]},
\label{eqb8}
\end{equation}
which can be simplified into
\begin{equation}
\langle U\rangle=-\frac{m\alpha^2}{\hbar^2}\left(\frac{1+2Ne^{-m\alpha a/\hbar^2}\sinh[m\alpha a/\hbar^2]}{1+2e^{-m\alpha a/\hbar^2}\left[\frac{1-(-1)^N}{2}\right]\sinh[m\alpha a/\hbar^2]}\right).
\label{eqb9}
\end{equation}
By using conservation of energy we can obtain the expectation value for the kinetic energy of the system.


\end{document}